\documentclass{elsart}
\usepackage{amssymb}

\renewcommand{\bar}[1]{\overline{#1}}

\usepackage{indentfirst}
\usepackage{psfig,color}
\usepackage{epsfig}
\usepackage{epsf}
\usepackage{graphicx}

\providecommand{\Journal}[4] {#1 {\bf #2} (#4) #3}
\providecommand{\PLB}{Phys. Lett. B } %
\providecommand{\PRL}{Phys. Rev. Lett. } %
\providecommand{\PRD}{Phys. Rev. D } %
\providecommand{\ZPC}{Z. Phys. C } %

\journal{Physics Letters B}

\begin{document}

\begin{frontmatter}
\title{A New Parametrization of the Neutrino Mixing Matrix}

\author[pku]{Nan Li},
\author[ccastpku]{Bo-Qiang Ma\corauthref{cor}}
\corauth[cor]{Corresponding author.} \ead{mabq@phy.pku.edu.cn}
\address[pku]{School of Physics, Peking University, Beijing 100871, China}
\address[ccastpku]{CCAST (World Laboratory), P.O.~Box 8730, Beijing
100080, China\\
School of Physics, Peking University, Beijing 100871, China}

\begin{abstract}
The neutrino mixing matrix is expanded in powers of a small
parameter $\lambda$, which approximately equals to 0.1. The
meaning of every order of the expansion is discussed respectively,
and the range of $\lambda$ is carefully calculated. We also
present some applications of this new parametrization, such as to
the expression of the Jarlskog parameter $J$, in which the
simplicities and advantages of this parametrization are shown.
\end{abstract}

\begin{keyword}
neutrinos \sep neutrino mixing matrix \sep parametrization \\
\PACS 14.60.Pq \sep 12.15.Ff \sep 13.15.+g \sep 14.60.Lm
\end{keyword}
\end{frontmatter}

\par

In recent years, there have been abundant experimental data
strongly suggesting the mixing of different generations of
neutrinos, just analogous to that of quarks. The K2K \cite{K2K}
and Super-Kamiokande \cite{SUPER} experiments indicated that the
atmospheric neutrino anomaly is due to the $\nu_{\mu}$ to
$\nu_{\tau}$ oscillation with almost the largest mixing angle of
$\theta_{atm} \approx 45^{\circ}$. The KamLAND \cite{Kam} and SNO
\cite{sno} experiments told us that the solar neutrino deficit was
caused by the oscillation from $\nu_{e}$ to a mixture of
$\nu_{\mu}$ and $\nu_{\tau}$ with a mixing angle approximately of
$\theta_{sol} \approx 34^{\circ}$. On the other hand, the
non-observation of the $\bar{\nu}_{e}$ to $\bar{\nu}_{e}$
oscillation in the CHOOZ \cite{Chz} experiment showed that the
mixing angle $\theta_{chz}$ is smaller than $3^{\circ}$ at the
best fit point \cite{ma,al}.

These experiments not only confirmed the oscillations of
neutrinos, but also measured the mass-squared differences of the
neutrino mass eigenstates (the allowed ranges at 3$\sigma$)
\cite{ma}, $1.4\times10^{-3}eV^{2}<\Delta
m_{atm}^{2}=|m_{3}^{2}-m_{2}^{2}|<3.3\times 10^{-3}eV^{2}$, and
$7.3\times10^{-5}eV^{2}<\Delta
m_{sol}^{2}=|m_{2}^{2}-m_{1}^{2}|<9.1\times 10^{-5}eV^{2}$, where
$\pm$ correspond to the normal and inverted schemes respectively.

Like the Cabibbo-Kobayashi-Maskawa (CKM) \cite{ckm1,ckm2} matrix
for quark mixing, the neutrino mixing matrix is described by the
unitary Maki-Nakawaga-Sakata (MNS) \cite{mns} matrix $V$, which
links the neutrino flavor eigenstates $\nu_{e}$, $\nu_{\mu}$,
$\nu_{\tau}$ to the mass eigenstates $\nu_{1}$, $\nu_{2}$,
$\nu_{3}$,
\begin{equation}
 \left(
    \begin {array}{c}
       \nu_{e} \\
       \nu_{\mu} \\
       \nu_{\tau} \\
\end{array}
\right) = \left(
    \begin {array}{ccc}
       V_{e1} & V_{e2} & V_{e3}\\
       V_{\mu1} & V_{\mu2} & V_{\mu3}\\
       V_{\tau1} & V_{\tau2} & V_{\tau3}\\
\end{array}
\right) \left(
    \begin {array}{c}
       \nu_{1} \\
       \nu_{2} \\
       \nu_{3} \\
\end{array}
\right).
\end{equation}

It is always feasible to parametrize the Majorana neutrino mixing
matrix as a product of a Dirac neutrino mixing matrix (with three
mixing angles and a CP-violating phase) and a diagonal phase
matrix (with three phase angles, and only two of them are
unremovable) \cite{sch}. In a form similar to the quark mixing
matrix, the neutrino mixing matrix can also be written as follows
\begin{equation}
    V=\left(
        \begin{array}{ccc}
            c_{2}c_{3} & c_{2}s_{3} & s_{2}e^{-i\delta}\\
           -c_{1}s_{3}-s_{1}s_{2}c_{3}e^{i\delta} & c_{1}c_{3}-s_{1}s_{2}s_{3}e^{i\delta} & s_{1}c_{2}\\
            s_{1}s_{3}-c_{1}s_{2}c_{3}e^{i\delta} & -s_{1}c_{3}-c_{1}s_{2}s_{3}e^{i\delta} & c_{1}c_{2}\\
        \end{array}
        \right)
 \left(
    \begin {array}{ccc}
     e^{i\sigma_{1}} &  & \\
       &e^{i\sigma_{2}}& \\
        &  & e^{i\sigma_{3}}\\
\end{array}
\right),
\end{equation}
where $s_{i}=\sin\theta_{i}$, $c_{i}=\cos\theta_{i}$ (for $i=1, 2,
3$), $\delta$ is the Dirac CP-violating phase and $\sigma_{1}$,
$\sigma_{2}$, $\sigma_{3}$ are the Majorana CP-violating phases.
If the neutrinos are of Dirac-type, the diagonal phase matrix on
the right side hand of Eq.~(2) can be rotated away by redefining
the phases of the Dirac neutrino fields. The Dirac CP-violating
phase is associated with the neutrino oscillations, CP and T
violation. The Majorana CP-violating phases are associated with
the neutrinoless double beta decay, and lepton-number-violating
processes \cite{sch1}.

The three mixing angles $\theta_{atm}$, $\theta_{chz}$, and
$\theta_{sol}$ are related to the three mixing angles
$\theta_{1}$, $\theta_{2}$, and $\theta_{3}$, which describe the
mixing between 2nd and 3rd, 3rd and 1st, 1st and 2nd generations
of neutrinos. To a good degree of accuracy,
$\theta_{atm}=\theta_{1}$, $\theta_{chz}=\theta_{2}$, and
$\theta_{sol}=\theta_{3}$.

According to the results of the global analysis of the neutrino
oscillation experiments, the elements of the modulus of the
neutrino mixing matrix are summarized as follows \cite{xing}
\begin{equation}
    |V|=\left(
        \begin{array}{ccc}
             0.70-0.87 & 0.50-0.69 & <0.16\\
             0.20-0.61 & 0.34-0.73 & 0.60-0.80\\
             0.21-0.63 & 0.36-0.74 & 0.58-0.80\\
        \end{array}
        \right).
\end{equation}

Quite different from the quark mixing matrix, almost all the
non-diagonal elements of the neutrino mixing matrix are large,
only with the exception of  $V_{e3}$. So it is unpractical to
expand the matrix in powers of one of the non-diagonal elements,
like the Wolfenstein parametrization of the quark mixing matrix
\cite{wol}. Xing \cite{xing1} has made the Wolfenstein-like
parametrization for the neutrino mixing matrix, but they have to
use much higher orders of the non-diagonal elements. In the quark
mixing pattern, all the non-diagonal elements are small, so we may
take it for granted that the mixing is a small modification to the
unit matrix. But on the contrary, why could not we consider the
large mixing as the common pattern, which is just the case in the
neutrino mixing? So we may not expand the neutrino mixing matrix
around the unit matrix.

In this letter, we will just make an expansion of the neutrino
mixing matrix based on the bi-maximal mixing pattern. Since there
are two mixing angles near $45^{\circ}$ ($\theta_{1} \approx
45^{\circ}$, and $\theta_{3} \approx 34^{\circ}$), the neutrino
mixing matrix is not only the bi-large pattern as commonly said,
but quite near the bi-maximal pattern, which reads
\begin{equation}
    V=\left(
        \begin{array}{ccc}
            \sqrt{2}/2 & \sqrt{2}/2 & 0 \\
            -1/2 & 1/2 & \sqrt{2}/2 \\
            1/2 & -1/2 & \sqrt{2}/2
        \end{array}
        \right).
\end{equation}

Comparing with Eq.~(3), we can make an expansion of $V$ in powers
of $\lambda$, which satisfies
\begin{equation}
  V_{e1}=\sqrt{2}/2+\lambda,
\end{equation}
where $\lambda$ measures the strength of the deviation of $V_{e1}$
from the bi-maximal mixing pattern. Unlike the Wolfenstein
parametrization of the quark mixing matrix, $\lambda$ here is at
the diagonal element of the neutrino mixing matrix. Because
$0.70<V_{e1}<0.87$, $\lambda$ is a small positive parameter, which
approximately equals to 0.1, and this expansion is reasonable and
will converge quickly. Because $\theta_{1}$ is quite near
$45^{\circ}$, $V_{\mu3}$ must be quite near $\sqrt{2}/2$
\cite{K2K,SUPER}. Then we can set
\begin{equation}
V_{\mu3}=\sqrt{2}/2+a\lambda^{2}.
\end{equation}
Also, since $\theta_{2}$ is rather small (with the best fit point
$V_{e3}=0.045$ \cite{ma}), we can set
\begin{equation}
V_{e3}=b\lambda^{2},
\end{equation}
where $a$ and $b$ are both small parameters of order 1.

Now we will calculate all the $s_{i}$ and $c_{i}$ (for $i=1, 2,
3$) to the order of $\lambda^{4}$. From Eq.~(7),
$s_{2}=V_{e3}=b\lambda^{2}$, we have
\begin{equation}
c_{2}=\sqrt{1-s_{2}^{2}}=1-\frac{1}{4}b^{2}\lambda^{4}.
\end{equation}
From Eq.~(6), we have
\begin{equation}
s_{1}c_{2}=V_{\mu3}=\sqrt{2}/2+a\lambda^{2},
\end{equation}
using Eq.~(8), we get
\begin{equation}
s_{1}=\frac{\sqrt{2}}{2}+a\lambda^{2}+\frac{\sqrt{2}}{4}b^{2}\lambda^{4}.
\end{equation}
Similarly
\begin{eqnarray}\nonumber
c_{1}&=&\frac{\sqrt{2}}{2}-a\lambda^{2}-\frac{1}{4}(4\sqrt{2}a^{2}+\sqrt{2}b^{2})\lambda^{4},\\\nonumber
c_{3}&=&\frac{\sqrt{2}}{2}+\lambda+\frac{\sqrt{2}}{4}b^{2}\lambda^{4},\\\nonumber
s_{3}&=&\frac{\sqrt{2}}{2}-\lambda-\sqrt{2}\lambda^{2}-2\lambda^{3}-\frac{1}{4}(12\sqrt{2}+\sqrt{2}b^{2})\lambda^{4}.\\
\end{eqnarray}
Thus we obtain all the trigonometric functions of the three mixing
angles.

Now we can get all the elements of the neutrino mixing matrix
straightforwardly,
\begin{eqnarray}\nonumber
V_{e1}&=&\frac{\sqrt{2}}{2}+\lambda,\nonumber\\
V_{e2}&=&\frac{\sqrt{2}}{2}-\lambda-\sqrt{2}\lambda^{2}-2\lambda^{3}-\frac{1}{2}(6\sqrt{2}+\sqrt{2}b^{2})\lambda^{4},\nonumber\\
V_{e3}&=&b\lambda^{2},\\
V_{\mu1}&=&-\frac{1}{2}+\frac{\sqrt{2}}{2}\lambda+\frac{1}{2}(2+\sqrt{2}a-b)\lambda^{2}+\frac{1}{2}(2\sqrt{2}-2a-\sqrt{2}b)\lambda^{3}\nonumber \\
&&+\frac{1}{2}(6-2\sqrt{2}a+2a^{2}-\sqrt{2}ab+b^{2})\lambda^{4},\nonumber\\
V_{\mu2}&=&\frac{1}{2}+\frac{\sqrt{2}}{2}\lambda-\frac{1}{2}(\sqrt{2}a+b)\lambda^{2}-\frac{1}{2}(2a-\sqrt{2}b)\lambda^{3}\nonumber \\
&&-\frac{1}{2}(2a^{2}-2b+\sqrt{2}ab)\lambda^{4},\nonumber\\
V_{\mu3}&=&\frac{\sqrt{2}}{2}+a\lambda^{2},\\
V_{\tau1}&=&\frac{1}{2}-\frac{\sqrt{2}}{2}\lambda-\frac{1}{2}(2-\sqrt{2}a+b)\lambda^{2}-\frac{1}{2}(2\sqrt{2}+2a+\sqrt{2}b)\lambda^{3}\nonumber \\
&&-\frac{1}{2}(6+2\sqrt{2}a-\sqrt{2}ab)\lambda^{4},\nonumber\\
V_{\tau2}&=&-\frac{1}{2}-\frac{\sqrt{2}}{2}\lambda-\frac{1}{2}(\sqrt{2}a+b)\lambda^{2}-\frac{1}{2}(2a-\sqrt{2}b)\lambda^{3}\nonumber \\
&&+\frac{1}{2}(2b+\sqrt{2}ab-b^{2})\lambda^{4},\nonumber\\
V_{\tau3}&=&\frac{\sqrt{2}}{2}-a\lambda^{2}-\frac{1}{2}(2\sqrt{2}a^{2}+\sqrt{2}b^{2})\lambda^{4}.
\end{eqnarray}

Then we can expand the neutrino mixing matrix in powers of
$\lambda$,
\begin{eqnarray}\nonumber
   V&=&\left(
        \begin{array}{ccc}
            \frac{\sqrt{2}}{2} & \frac{\sqrt{2}}{2} & 0 \\
            -\frac{1}{2} & \frac{1}{2} & \frac{\sqrt{2}}{2} \\
            \frac{1}{2} & -\frac{1}{2} & \frac{\sqrt{2}}{2}
        \end{array} \right)+\lambda
 \left(
        \begin{array}{ccc}
            1 & -1 & 0 \\
            \frac{\sqrt{2}}{2} & \frac{\sqrt{2}}{2}  & 0\\
           -\frac{\sqrt{2}}{2} & -\frac{\sqrt{2}}{2} & 0
        \end{array} \right)+\lambda^{2}
\left(
        \begin{array}{ccc}
            0 & -\sqrt{2} & b \\
            \frac{1}{2}(2+\sqrt{2}a-b) & -\frac{1}{2}(\sqrt{2}a+b) & a \\
            -\frac{1}{2}(2-\sqrt{2}a+b) & -\frac{1}{2}(\sqrt{2}a+b) & -a
        \end{array} \right)\\&&+\lambda^{3}
 \left(
        \begin{array}{ccc}
            0 & -2 & 0 \\
            \frac{1}{2}(2\sqrt{2}-2a-\sqrt{2}b) & -\frac{1}{2}(2a-\sqrt{2}b)  & 0\\
           -\frac{1}{2}(2\sqrt{2}+2a+\sqrt{2}b) & -\frac{1}{2}(2a-\sqrt{2}b)  & 0
        \end{array} \right)+\cdots.
\end{eqnarray}

Now we will see the meaning of every order in the expansion of
$V$.

1. The term of $\lambda^{0}$ is the approximation of the lowest
order, where the atmospheric and solar neutrino oscillations are
both of the largest mixing angles of $45^{\circ}$. We call this
the bi-maximal mixing pattern.

2. The term of $\lambda^{1}$ indicates the deviation of the
neutrino mixing matrix from the bi-maximal mixing pattern.

3. The term of $\lambda^{2}$ shows the effect of the CP violation.
Because the CP violation is described by the element $V_{e3}$
\cite{prdd}, and in the terms of $\lambda^{0}$ and $\lambda^{1}$,
$V_{e3}=0$, the degree of the CP violation is of the order
$\lambda^{2}$ in our parametrization.

4. The term of $\lambda^{3}$ is the modification of higher order.

So in our new parametrization the terms of $\lambda^{0}$,
$\lambda^{1}$, $\lambda^{2}$ show the bi-maximal mixing pattern,
the deviation from the bi-maximal mixing pattern, and the CP
violation effect respectively.

Next, we are going to determine the ranges of $\lambda$, $a$ and
$b$. From the analysis above, we know that
\begin{eqnarray}\nonumber
\sin\theta_{atm}&=&s_{1}=\frac{\sqrt{2}}{2}+a\lambda^{2}+\frac{\sqrt{2}}{4}b^{2}\lambda^{4},\nonumber\\
\sin\theta_{chz}&=&s_{2}=b\lambda^{2},\nonumber\\
\sin\theta_{sol}&=&s_{3}=\frac{\sqrt{2}}{2}-\lambda-\sqrt{2}\lambda^{2}-2\lambda^{3}-\frac{1}{4}(12\sqrt{2}+\sqrt{2}b^{2})\lambda^{4}.
\end{eqnarray}

The current experimental data of these three parameters are (the
allowed ranges at 3$\sigma$) \cite{ma,al}
\begin{eqnarray}\nonumber
0.58&<\sin\theta_{atm}<&0.81,\nonumber\\
0&<\sin\theta_{chz}<&0.16,\nonumber\\
0.48&<\sin\theta_{sol}<&0.61.
\end{eqnarray}

And the best fit points are \cite{ma,al}
\begin{eqnarray}\nonumber
\sin\theta_{atm}&=&0.72,\nonumber\\
\sin\theta_{chz}&=&0.045,\nonumber\\
\sin\theta_{sol}&=&0.55.
\end{eqnarray}

From these three constraints, we can determine the ranges of the
three parameters $\lambda$, $a$ and $b$. First, we can determine
the range of $\lambda$. From Eq.~(16), we have
\begin{eqnarray}\nonumber
0.48<\frac{\sqrt{2}}{2}-\lambda-\sqrt{2}\lambda^{2}-2\lambda^{3}-\frac{1}{4}(12\sqrt{2}+\sqrt{2}b^{2})\lambda^{4}<0.61,
\end{eqnarray}
then

\begin{eqnarray}\nonumber
\frac{0.28-2.83\lambda-4\lambda^{2}-5.66\lambda^{3}-12\lambda^{4}}{\lambda^{4}}
<b^{2}<
\frac{0.65-2.83\lambda-4\lambda^{2}-5.66\lambda^{3}-12\lambda^{4}}{\lambda^{4}}.
\end{eqnarray}

From the right part of the inequality, we have
\begin{eqnarray}
0.65-2.83\lambda-4\lambda^{2}-5.66\lambda^{3}-12\lambda^{4}>0.
\end{eqnarray}
The allowed range of $\lambda$ is shown in Fig.~1, and we can see
that $-0.67<\lambda<0.17$.

From the left part of the inequality, if
\begin{eqnarray}\nonumber
0.28-2.83\lambda-4\lambda^{2}-5.66\lambda^{3}-12\lambda^{4}>0,
\end{eqnarray}
we have $-0.62<\lambda<0.08$, which does not agree with the value
of $V_{e1}$. So we must set
\begin{eqnarray}
0.28-2.83\lambda-4\lambda^{2}-5.66\lambda^{3}-12\lambda^{4}<0,
\end{eqnarray}
and thus the inequality satisfies automatically. The allowed range
of $\lambda$ is shown in Fig.~1, and we can see that
$\lambda<-0.62$ or $\lambda>0.08$. Summarizing these results, we
get $0.08<\lambda<0.17$, which is consistent with the primary
estimation in Eq.~(5). So it is reasonable and practical to expand
the neutrino mixing matrix in powers of $\lambda$.

\begin{figure}
\begin{center}
\includegraphics[width=7cm]{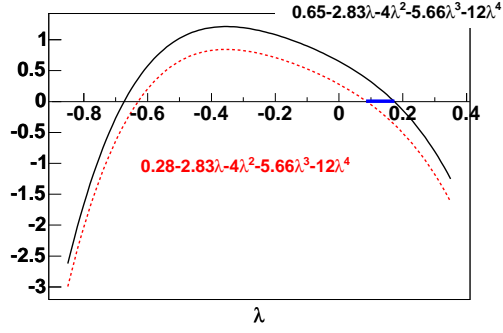}
\end{center}
\caption{The allowed range of $\lambda$, the solid curve is for
$0.65-2.83\lambda-4\lambda^{2}-5.66\lambda^{3}-12\lambda^{4}=0$,
and the dashed curve is for
$0.28-2.83\lambda-4\lambda^{2}-5.66\lambda^{3}-12\lambda^{4}=0$.}\label{fig1}
\end{figure}

Now we can determine the range of $b$. Because
$b\lambda^{2}=\sin\theta_{2}=\sin\theta_{chz}$, using Eqs.~(17),
(18) and $0.08<\lambda<0.17$, we have that $b=0.045/\lambda^{2}$.
The range of $b$ is shown in Fig.~2, and we can see that
$1.56<b<7.03$.

\begin{figure}
\begin{center}
\includegraphics[width=7cm]{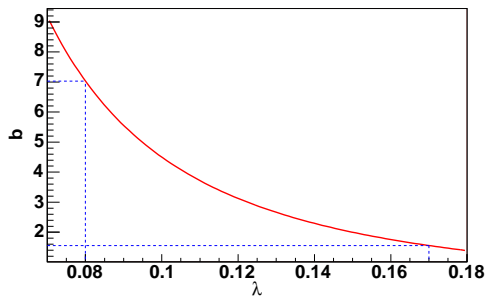}
\end{center}
\caption{The range of $b$.}\label{fig2}
\end{figure}

Similarly, in the case of $a$,
$V_{\mu3}=\sqrt{2}/2+a\lambda^{2}=s_{1}c_{2}$. Using Eq.~(17) and
Eq.~(18), we have $0.58<s_{1}c_{2}=s_{1}\sqrt{1-s_{2}^{2}}<0.81$,
with the best fit point 0.72. Thus $\sqrt{2}/2+a\lambda^{2}=0.72$,
so $a=0.01/\lambda^{2}$. The range of $a$ is shown in Fig.~3, and
we can see that $0.35<a<1.56$.

\begin{figure}
\begin{center}
\includegraphics[width=7cm]{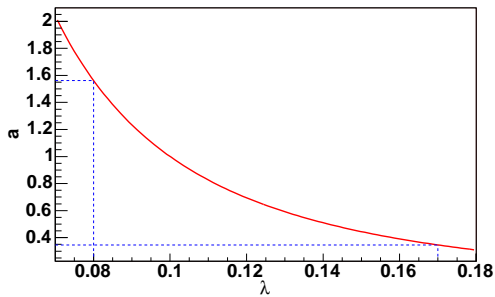}
\end{center}
\caption{The range of $a$.}\label{fig3}
\end{figure}

In our new parametrization, several other corresponding observable
quantities associated with the elements of the neutrino mixing
matrix can be expressed in relatively simple forms. From the
ranges of $\lambda$, $a$ and $b$, we can determine the ranges of
these observable quantities.

1. The Jarlskog parameter $J$ \cite{Ja}. $J$ is the
rephasing-invariant measurement of the lepton CP violation. The
Majorana  CP-violating phases can be removed away by redefining
the phases of the Dirac fields, so only $\delta$ is associated
with the CP violation.
$J=\mbox{Im}(V_{e2}V_{\mu3}V_{e3}^{\ast}V_{\mu2}^{\ast})=s_{1}s_{2}s_{3}c_{1}c_{2}^2c_{3}\sin\delta$.
In our parametrization, $J$ can be expressed in a very simple form
(to the order of $\lambda^{4}$)
\begin{equation}
J=(\frac{\sqrt{2}}{2})^{4} b\lambda^{2}\sin\delta(1-4\lambda^{2}).
\end{equation}
Because $s_{1}$, $s_{3}$, $c_{1}$, $c_{3}$ all have the factor
$\sqrt{2}/2$, there are four $\sqrt{2}/2$ in $J$. So the degree of
the lepton CP violation is suppressed four times. Again, $J$  is
suppressed by the factor $b\lambda^{2}=0.045$ \cite{ma}. Using
$0.08<\lambda<0.17$, $1.56<b<7.03$, we can determine the range of
$J$ in Fig.~4, and we can see that $J\approx0.00996\sim 0.01096$
(here we take $\sin\delta\sim1$).

\begin{figure}
\begin{center}
\includegraphics[width=7cm]{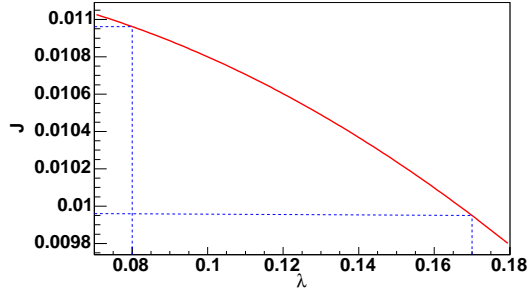}
\end{center}
\caption{The range of $J$.}\label{fig4}
\end{figure}

2. The effective Majorana mass term $\langle m \rangle_{ee}$. In
the neutrinoless double beta decay, the effective Majorana mass
term is defined as follows
\begin{eqnarray}\nonumber
\langle
m\rangle_{ee}\equiv|m_{1}V_{e1}^{2}e^{2i\sigma_{1}}+m_{2}V_{e2}^{2}e^{2i\sigma_{2}}+m_{3}V_{e3}^{2}e^{2i\sigma_{3}}|,
\end{eqnarray}
where $\sigma_{1}$, $\sigma_{2}$, $\sigma_{3}$ are the Majorana
CP-violating phases \cite{sch1}. Using Eq.~(12), we get
\begin{eqnarray}
\langle m
\rangle_{ee}=&|\frac{1}{2}(m_{1}e^{2i\sigma_{1}}+m_{2}e^{2i\sigma_{2}})+(\sqrt{2}\lambda+\lambda^{2})(m_{1}e^{2i\sigma_{1}}
-m_{2}e^{2i\sigma_{2}}) \nonumber \\
&+b^{2}\lambda^{4}(m_{3}e^{2i\sigma_{3}}-m_{2}e^{2i\sigma_{2}})|.
\end{eqnarray}
We can see that the coefficients of the three terms show the
influences of the three orders of $\lambda$. Only $m_{1}$ and
$m_{2}$ are important to the value of $\langle m \rangle_{ee}$,
and influence of $m_{3}$ almost vanish if the masses of the three
mass eigenstates are nearly degenerated, because the coefficient
$b^{2}\lambda^{4}$ is of $10^{-3}$.

3. The effective mass terms of neutrinos. The effective mass terms
of neutrinos can be defined as follows (here we take electron
neutrino for example.)
\begin{eqnarray}\nonumber
\langle m\rangle_{e}^{2}\equiv
m_{1}^{2}|V_{e1}|^{2}+m_{2}^{2}|V_{e2}|^{2}+m_{3}^{2}|V_{e3}|^{2}.
\end{eqnarray}
Using Eq.~(12), we get
\begin{equation}
\langle m\rangle_{e}^{2}
=\frac{1}{2}(m_{1}^{2}+m_{2}^{2})-(\sqrt{2}\lambda+\lambda^{2})(m_{2}^{2}-m_{1}^{2})+
b^{2}\lambda^{4}(m_{3}^{2}-m_{2}^{2}).
\end{equation}
Again, the coefficients of the three terms show the influences of
the three orders of $\lambda$. Noting that $\Delta
m_{sol}^{2}=|m_{2}^{2}-m_{1}^{2}|$ and $\Delta
m_{atm}^{2}=|m_{3}^{2}-m_{2}^{2}|$, we can rewrite Eq.~(21) into
\begin{equation}
\langle m \rangle_{e}^{2}=m_{1}^{2} \pm
[\frac{1}{2}-(\sqrt{2}\lambda+\lambda^{2})]\Delta m_{sol}^{2} \pm
b^{2}\lambda^{4}\Delta m_{atm}^{2}.
\end{equation}
We can see from Eq.~(24) that $\langle m \rangle_{e}^{2}$ is
directly related with the masses and the mass-squared differences
of neutrinos. So these two kinds of different observable
quantities are associated together in our parametrization. If we
can separately measure $\Delta m_{atm}^{2}$, $\Delta m_{sol}^{2}$,
and $\langle m \rangle_{e}^{2}$ to a good degree of accuracy, we
can fix the value of $m_{1}$, which will help us determine the
absolute mass of neutrino ultimately.

In summary, although all kinds of parametrization of the neutrino
mixing matrix are mathematically equivalent, and applying any of
them does not have any specific physical significance, however, it
is quite likely that some particular parametrization does have its
usefulness and advantages in analysis of various experimental
data. Furthermore, we can express other observable quantities in a
simple and transparent way, and can link several different kinds
of observable quantities together. This is the purpose of our new
parametrization, and we hope that this new parametrization will be
useful in the phenomenology of neutrino physics.

{\bf Acknowledgments }

We are grateful for the discussions with Prof. Zhizhong Xing. This
work is partially supported by National Natural Science Foundation
of China under Grant Numbers 10025523 and 90103007.

\end{document}